\documentclass{article}

\usepackage{arxiv}

\usepackage{graphicx,url}
\usepackage[utf8]{inputenc}
\usepackage[english]{babel}

\title{Developing Algorithms for the Internet of Flying Things Through Environments With Varying Degrees of Realism - Extended Version}

\author{
Thiago de Souza Lamenza \\
  Department of Informatics\\
  Pontifícia Universidade Católica do Rio de Janeiro\\
  Rio de Janeiro, Brazil \\
  \texttt{tlamenza@inf.puc-rio.br} \\
  \And
 Josef Kamysek \\
  Department of Informatics\\
  Pontifícia Universidade Católica do Rio de Janeiro\\
  Rio de Janeiro, Brazil \\
  \texttt{josef@kamysek.com} \\
   \And
 Bruno José Olivieri de Souza \\
  Department of Informatics\\
  Pontifícia Universidade Católica do Rio de Janeiro\\
  Rio de Janeiro, Brazil \\
  \texttt{bolivieri@inf.puc-rio.br} \\
  \And
 Markus Endler \\
  Department of Informatics\\
  Pontifícia Universidade Católica do Rio de Janeiro\\
  Rio de Janeiro, Brazil \\
  \texttt{endler@inf.puc-rio.br} \\
}

\begin{document} 

\maketitle

\begin{abstract}
This work discusses the benefits of having multiple simulated environments with different degrees of realism for the development of algorithms in scenarios populated by autonomous nodes capable of communication and mobility. This approach aids the development experience and generates robust algorithms. It also proposes GrADyS-SIM NextGen as a solution that enables development on a single programming language and toolset over multiple environments with varying levels of realism. Finally, we illustrate the usefulness of this approach with a toy problem that makes use of the simulation framework, taking advantage of the proposed environments to iteratively develop a robust solution.
\end{abstract}

\section{Introduction}
When working with the development of distributed systems populated by autonomous nodes capable of movement, you end up dealing with networking and mobility, which highly affect the algorithm's performance and behavior \cite{bruno-experiments}. Algorithms running in these environments require specialized features to achieve a desired level of robustness. The importance of these features doesn't become apparent until the algorithm is exposed to the conditions that motivate them. The process necessary to create and validate these systems very time-consuming and monetarily draining when reliant on field tests. For these reasons, using a simulated environment to aid the research and creation processes is essential. Simulation is a common approach to implementing solutions to problems in these scenarios \cite{UAVApplication} \cite{FlyingAdHocNetworks} \cite{SurveyonUnmannedVehicles}.

Representing the real world in a simulated environment is very important as it exposes the algorithms to scenarios and challenges it will encounter in a final deployment without incurring the costs of field tests. Fidelity is key here, the robustness of an algorithm in development inside a simulation is directly proportional to the amount of real world conditions that can be thrown at it, motivating the addition of features that bypass issues that arise from the conditions it was exposed to. 

One of the ways to tackle the development of these algorithms is through an empirical and iterative process of development and evaluation. The concept of incremental and iterative development is not new and is popular in software engineering in general. Representing and simulating too many specific aspects of the real world at the same time hinders the development process, which should start simple in an environment with most of it abstracted away and gradually introduce realism until the desired requirement of robustness is achieved. The increment in realism comes at a cost because with the increase in complexity of the simulation software, the development experience tends to deteriorate. A realistic simulation has a heavy computational cost that adds a bigger overhead to the execution of simulation scenarios, slowing down the development process. The complexity of the software also comes with other disadvantages as they tend to be harder to set up, have a steeper learning curve and be very specialized, making it hard to write code that translates well to other environment and the real world. 

This work presents \emph{GrADyS-SIM NextGen}, a framework for simulating distributed algorithms in a simulated network environment populated by nodes capable of communication and mobility. A short video showcase of the simulator is available \footnote{\url{https://youtu.be/vixp_CicbJk}}.There are ample use cases for a simulator like this. Examples are simulating systems where unmanned aerial vehicles communicate with stationary sensors, the simulation of UAV formations that rely on communication to maintain a desired configuration, predator and prey scenarios and many more. The simulator presented in this work is not unique in the area of simulating networks of UAVs \cite{flynetsim} and work has been done in compiling comprehensive lists of existing options \cite{SurveyNetworkSimulators} \cite{UAVSimulators}. GrADyS-SIM NextGen distinguishes itself from the rest for its philosophy of encouraging the iterative and empirical approach proposed in this work.

That main distinguishing factor manifests itself with the proposal of a general and environment agnostic way of implementing distributed algorithms that can run in different simulated environments and, in principle, in the real world. This has several benefits, the main of which are the lowering of the learning curve for the framework since only one interface has to be learned for algorithm development and the removal of the effort required to translate code between different environments with different requirements. Having a common interface also enables the creation of tooling that improves the experience of creating distributed algorithms.

The framework enables you to create artifacts called \emph{protocols} which, using Python, implement the logic that powers individual nodes. These Python protocols can be used in a simple Python simulation environment, which we call \emph{prototype-mode}. Users can also run them in \emph{integrated-mode} which will integrate with OMNeT++, an event-based network simulator, to simulate realistic network interactions and even integrate it with SITL through the use of MAVSIMNET \footnote{\url{https://github.com/Thlamz/MAVSIMNET}} which provides a realistic mobility models representing several vehicle types. Finally, these protocols can be used to control real-world vehicles in what we call \emph{experiment-mode}, although this last execution mode is currently not implemented yet. The main appeal of the framework is that no code changes need to be made to the protocol itself when switching between these modes, making the knowledge obtained in a simulation environment easily transferable to another or to the real world. 

This is a continuation of \emph{GrADyS-SIM} \cite{gradysim} with which it shares its main objectives. The creation of this new simulation framework based on GrADyS-SIM was motivated by our experience in previous works, feedback from both within the team and from external users, and the realization that our current framework at the time made translating our simulated implementations to real-life for experiments very hard.

The software was created as part of the GrADyS \cite{gradys2021} (Ground-and-Air Dynamic sensor networkS) project. One of the project's main interests is investigating the use of autonomous aerial vehicles in data collection scenarios, such as remote or dangerous areas not easily accessible by people. GrADyS-SIM was developed as a simulation tool for the project first and foremost, its general usage was limited. A more general and user-focused approach is being taken with GrADyS-SIM NextGen without losing its utility to the project.

This paper is organized as follows. In section \ref{motivation} we will discuss the concepts that guided the framework's implementation. Section \ref{architecture} the framework's components will be presented in a lower level. In section \ref{results_and_discussions} a demonstration of the usage of GrADyS-SIM NextGen is presented as empirical evidence that the proposed development approach has tangible benefits for algorithm development.

\section{Motivation} \label{motivation}

The GrADyS-SIM simulator was used in several projects internal to the GrADyS team and by some external users. A common pain point observed by both of these groups was the difficulty of setting up and using OMNeT++ \footnote{\url{https://omnetpp.org/}} and its component library INET \footnote{\url{https://inet.omnetpp.org/}}. OMNeT++ is absolutely essential to the project for its network simulation capabilities but using it comes with a couple of downsides. It is a very large piece of software, even the IDE it comes packaged with is slow even on modern machines, build times and even simple tasks like checking a function's usages in code take a while making development inconvenient. Its complex structure and ample set of features means that it has a steep learning curve and many possible points of failure. These failures are not always easy to debug, in fact a significant portion of the simulator user's effort is spent doing exactly that. To top it off, it is not trivial to set up on modern machines, often leading to users ending up with broken or incomplete installations. These factors combined discourage potential users, even ones internal to the team.

Dropping support for OMNeT++ was not an option. As stated, its features are essential to running realistic simulations in the project. Having it as a requirement was also undesired because as long as people have to use it, the bar of entry for our simulation framework will always be high. A solution was devised that satisfies both requirements adequately. OMNeT++ would no longer be a requirement to run this new version of the simulation framework. Instead, it would be an optional dependency if the user desired a realistic network simulation. How this would be achieved is related to another one of this work's guiding concepts.

GrADyS-SIM is based on OMNeT++ which uses C++ as its implementation language. The project's real world testbed uses Python to implement node behavior. This means that anything developed inside the simulation would then need to be translated into Python, potentially introducing errors and requiring more development time. This led to the choice of Python as the language in which protocol logic would be written in. It is a convenient language with a low learning curve, which is both attractive to newcomers and compatible with our testbed. It's also easy to integrate in C/C++ code, making interoperability with OMNeT++ achievable.

Having a single language to define the algorithms doesn't come close to solving all problems. First, there is no guarantee that the same code will work on all the required environments, with their own set of requirements. Also, having protocols written in Python doesn't make the simulation framework much easier to use, as users will still need to deal with OMNeT++ as the simulation software. Two more decisions were made to guarantee the framework would meet its requirements, an environment agnostic interface for defining protocols was needed and a simpler simulation environment for cases where OMNeT++ wasn't necessary.

The protocol interface was created to serve as an environment-agnostic interface to implement algorithms. Code that uses it should be able to run in any environment supported by the framework, be it simulated or real. It establishes a well-defined set of rules a protocol should follow in order to be compliant with the interface and enjoy the benefits of being decoupled from their environment. The main benefit is reducing or even removing the effort required to adapt them to new environments. This reduces the overhead required when moving tests between simulated environments and the real world. 

Reduced overhead is not the only benefit of following the protocol interface. Having a generalized API that protocols adhere to enable the creation of a toolset dedicated to aiding the creation of new protocols. Repetitive boilerplate work is required when creating new protocols and has to be re-implemented every time a new protocol needs to use it. Without a standard interface, protocol code is not easily reusable. It's not that creating tools to reduce this repetitive work isn't possible, but the incentive to do so isn't big because any code would need to be adapted when the simulation's characteristics changed. With a generic interface established, tools no longer need to be adapted to new situations because they use the same features available to the protocols themselves through the protocol interface. A set of tools is provided with the simulator for common tasks like implementing leader-follower mobility, random mobility and a mobility that follows a mission. The fact that most of the plugins are related to mobility are merely coincidental, the possible use cases are far wider.

The last of the requirements to be fulfilled is reducing the barrier of entry by making the simulation framework easier to install, understand and experiment with. This was achieved by creating a new environment, protocols running in this mode are said to be running in \emph{prototype-mode}. This environment consists of a Python event-based simulator created specifically for the simulation framework. It is trivial to install, light on dependencies and works on Python version 3.8 or higher. 

Python's simple syntax combined with the simplicity of the simulator provides a good starting point for people unfamiliar not only with the project but with how an event-based simulator works and how it can assist you in implementing network simulations. This comes at the cost of fidelity. The prototype-mode simulator lacks almost all the features that OMNeT++ has, providing only the absolute basics. The intention of prototype-mode is being an accessible and simple environment where simulations are quick to set up and run. Development is an iterative process, and having an environment that reduces the overhead of running experiments is a great luxury. Also, when users are in the early stages of developing a new algorithm, having access to a realistic network and mobility simulation is not very important, testing the functionality of a protocol on a simpler environment is more than adequate. 

When users are done with their prototypes, they can still use OMNeT++ to perform realistic simulations, just like the last version of the framework. The same exact protocol code can be used on both environments, prototype-mode and integrated-mode. Work was also employed into making the setup of OMNeT++ easier. It is still the harder of the two simulation environments to set up, specially as the python integration adds some complexity to the setup. The docker container provided in the repository's documentation should at least alleviate the pain of setting it up. 

\section{Architecture} \label{architecture}

\begin{figure}
    \centering
    \includegraphics[width=0.73\linewidth]{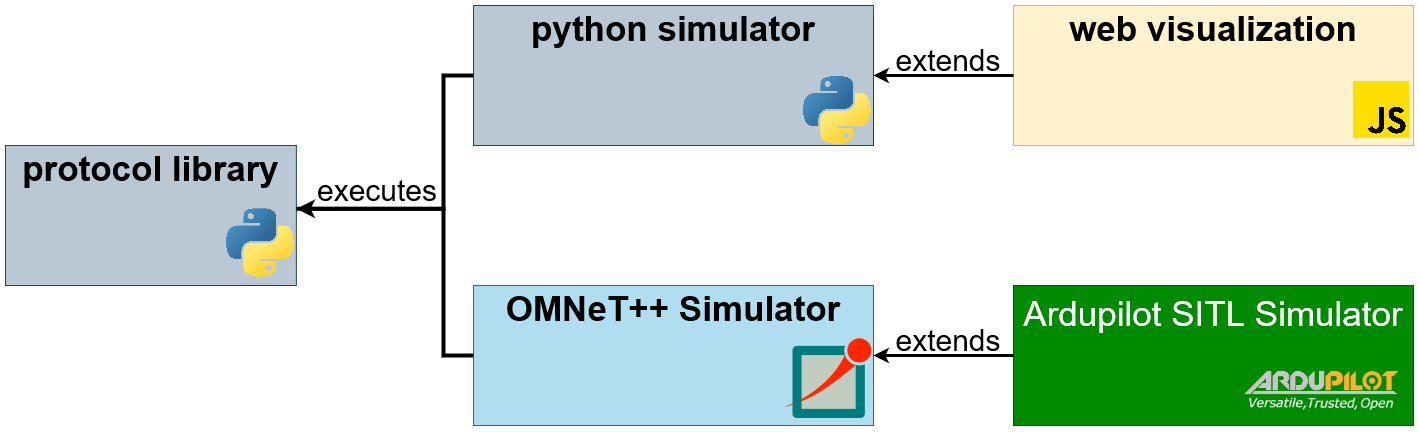}
    \caption{Framework's architecture}
    \label{fig:architecture}
\end{figure}

GrADyS-SIM NextGen is a simulation framework with several components. All components are publicly available and open-source. The framework is distributed in three GitHub repositories, \emph{gradys-sim-nextgen} \footnote{\url{https://github.com/Project-GrADyS/gradys-sim-nextgen}} contains the python components of the framework including the protocol library and the python simulator, \emph{gradys-sim-visualization} \footnote{\url{https://github.com/Project-GrADyS/gradys-sim-nextgen-visualization}} hosts the visualization website for python simulations and \emph{gradys-simulations} \footnote{\url{https://github.com/Project-GrADyS/gradys-simulations}} hosts the OMNeT++ components of the framework and the integration with Ardupilot's SITL \footnote{\url{https://ardupilot.org/}}. The documentation for each of these components is also hosted in their own repositories.

There are three main ways of running simulations in the framework. The first one is implementing your algorithms using the protocol library and running them in the python simulator. Using this scheme, you will have access to all protocol features while running in a simplified environment with low implementation overhead and fast execution times. It is ideal for prototyping and quick development. Optionally, you can use the web visualizer when running simulations like this. When looking for more serious results using the same code base, one can try integrating with OMNeT++ and running your code built on the protocol library in the OMNeT++ simulated environment. Lastly, building simulations completely in OMNeT++ is still supported, you will be creating all your code in C++ and won't be able to use it elsewhere. Both options that use OMNeT++ can optionally integrate with the Ardupilot SITL Simulator for a better mobility model.

In the rest of this section, we will talk about each of the individual components seen in figure \ref{fig:architecture}, detailing their purpose in the framework and how they are distributed.

\subsection{Protocol library}

\begin{figure}
    \centering
    \includegraphics[width=0.55\linewidth]{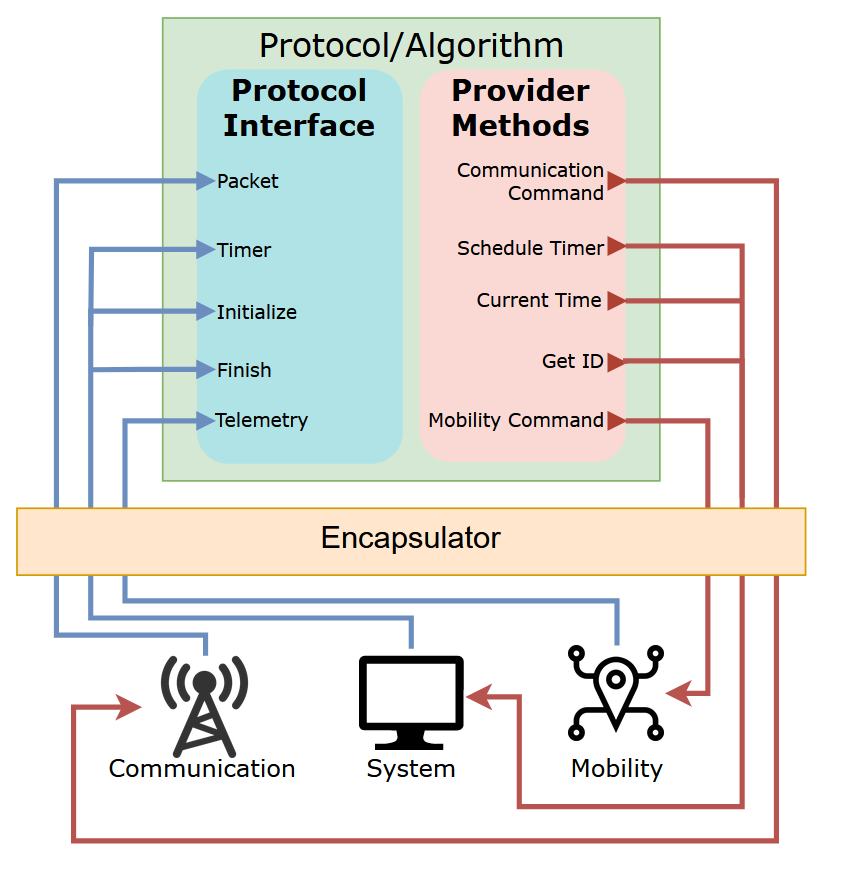}
    \caption{Diagram showcasing how protocols work}
    \label{fig:protocol-diagram}
\end{figure}

To allow distributed algorithm logic, implemented as protocol code, to run in supported environments, be it simulated or real, a common and general interface needed to be established defining how information from the environment would be available to the protocol and how it interacts with it. The environment's influence in the protocol was modeled as an interface called protocol interface. The protocol's interactions with the environment are defined through an interface called a \emph{provider interface}. This library contains the code necessary to allow users to create their own protocols. A protocol must strictly follow the template provided by the protocol interface in order to properly function in the supported environments. In concrete terms this means implementing all the required methods and only using functions provided by the provider interface to interact with the environment.

Protocols are implemented in an event-oriented way. The events available to the protocol were chosen carefully as to give them the necessary information to implement their behavior, but not couple them to a specific environment. Since protocols power nodes capable of communication and mobility, they were two of the main focuses of the interface. Protocols are able to react to messages received, being able to read their contents. They are also constantly updated about their mobility status, mainly about their position. The other methods are related to more basic needs. Initialization and finalization events are provided as starting and ending point for protocol logic. Distributed algorithms often depend on performing actions at certain times or intervals, so protocols are made aware of when their self-defined timers fire. 

It's also essential that the protocol can act upon the environment. These actions are performed with the provider interface, named this way because it \emph{provides} the protocol with the means to act in its environment. The tools available to the protocol are highly dependent on the environment, if we want to build an environment agnostic piece of code we can't deal with details like the physical means of locomotion of the node or the hardware that makes it capable of communication. For this reason, a generic set of messages were created to hide these details from the protocol through a layer of abstraction.

Mobility commands are messages that define the primitives through which the protocol will control its mobility. This simple set of primitives allows the protocol to go to a coordinate, geographical or Euclidean, and set its speed. Much more complex behavior can be built on top of this simple set of commands. Communication commands define the actions the protocol is capable of performing to impact its communication. It is also a very simple set of primitives that allow it to send targeted or broadcast messages. The complexities of the real or simulated environment where the node is in are thus hidden from it behind these abstractions.

Commands are sent by the protocol through the provider interface. The actual implementation of the provider interface is environment-dependent, it is injected into the protocols at run-time and the protocol does not rely on its design, only that it follows the specification defined in the interface. Communication commands may result into calls to transmission hardware to send a message, or send a message through a simulated network stack inside OMNeT++. The protocol is not concerned with how its commands are handled.

The common interfaces that define a protocol allow us to run the same code in any environment, provided that some glue code called an \emph{encapsulator} is written to adapt the protocol and provider interfaces to the environment. This glue code is written on a per-environment basis, not a per-protocol basis, so the effort to maintain it is not excessive and a normal user shouldn't need to write any unless he is intending to support some new environment. 

Another feature available in the protocol library is a set of tools named plugins. They come with varying objectives, but mainly focused on making implementing new protocols easier. They range from abstracting common behaviors in distributed algorithms from facilitating the implementation of movement patterns. These tools are only exposed to the protocol interface, and thus protocols that use them are still completely compatible with it.

This work proposes that the defined library is adequately capable of representing a wide array of use cases in the creation of distributed algorithms, in section \ref{demonstration} we will demonstrate this premise by giving a concrete use case.

Figure \ref{fig:protocol-diagram} illustrates how all of these ideas fit together. In the figure, the protocol interface and provider methods are shown to be separated from the environment by the encapsulator. The environment is represented in high abstraction by three components, mobility, communication, and system. \emph{Mobility} is an abstraction for whatever empowers nodes with mobility, be that hardware or simulated code. Likewise, \emph{communication} represents what gives the node the power to communicate. Lastly, \emph{system} represents a component that provides computation context and tasks, this could be the operating system in a real scenario or the simulator itself in a simulated scenario.

Everything described in this section is available in the \emph{gradys-sim-nextgen} repository and can be installed directly from PyPI, Python's package repository under the name \emph{gradysim}.

\subsection{Python simulator}

The simulator module is a Python package that implements an event-based network simulator. It was designed to be of similar usage as OMNeT++ but with a much simpler interface, gentle learning curve, trivial installation process and a limited set of features. The main objective of this simulator is serving as a platform for prototyping and developing protocols. Its low overhead and streamlined usage allow users to quickly iterate through versions of their protocols, improving developer experience.

Although simple in terms of functionality, it is still capable of emulating a lot of circumstances a real node will be exposed to. It simulates communication, allowing users to specify communication range and introducing common network failures such as network delay or dropped messages. It also simulates mobility, which is essential since this entire framework has been created to allow for the simulation of networks populated by mobile nodes. 

This component fits in to the framework as a first step for protocol development. The simpler simulator also serves as a great entry-point for new users who can learn how to create and implement distributed algorithms. The Python simulator is also part of the \emph{gradysim} Python package, and its source code is available in the \emph{gradys-sim-nexgen} repository.

\subsection{Web visualization}
This component integrates with the python simulator to provide a visual representation of a running simulation. Nodes are displayed as spheres positioned in a 3D visual environment, where the ground is marked as a black mesh. Information about the running simulation like the simulation time and the value of some variables marked for tracking inside protocols can be seen in the user interface. Users can also color specific nodes to distinguish them from the rest, this is very important to visually interpret some simulations. The component is available as a website \footnote{\url{https://project-gradys.github.io/gradys-sim-nextgen-visualization/}}. 

The visualization works by consuming information from a Web Socket. When the simulation starts, the visualization component of the python simulator, which is optional, opens a Web Socket server locally. On detecting that this server is online, the visualization website connects to it and starts receiving live information from the simulator.

This strategy was chosen because the Python libraries capable of 3D rendering that were sampled were not fitting with the project's requirement of simplicity, ease of setup and minimal dependencies. Although complicated at a first glance, the visualization was implemented in less than 300 lines of Typescript code and has only one dependency for the 3D rendering, \emph{Three.JS} \footnote{\url{https://threejs.org/}}. Also, the choice of implementing it on the web means there are zero installation steps if the user already has a web browser installed.

\subsection{OMNeT++}

The next big component is the OMNeT++ simulator. This whole component was already available in the previous version of the simulation framework. It has been modified slightly to account for interacting with the encapsulator module. The previous version of the simulator was already built on the idea of a central protocol module implementing the node's behavior, so adapting was easy. To bridge the communication gap between Python and OMNeT++, a proxy protocol sits in the C++ code acting as a protocol but redirecting all calls to the Python protocol. It acts as a middleman, ensuring that commands and information flow between the two environments. 

Users can select the Python implemented protocol they want through OMNeT++'s configuration system. In execution time, this protocol will be imported and wrapped with an \emph{encapsulator}. The proxy protocol will redirect commands to this wrapped protocol and execute any actions it commands. Protocols running in this way are said to be running in \emph{integrated-mode}. All interactions with python are managed through pybind11 \footnote{\url{https://github.com/pybind/pybind11}}.

\subsection{Ardupilot SITL Simulator}

The final component in the architecture is entirely optional. The OMNeT++ simulation can integrate with Ardupilot's SITL Simulator. The SITL simulator allows users to simulate real vehicles running Ardupilot with a convincing mobility model that takes into account the physics of the vehicle and the environment. Through work developed in MAVSIMNET, integrating SITL into OMNeT++ is possible. This provides a much more realistic mobility model to the simulated node's movement. The integration replaces the mobility model implemented as part of the framework's OMNeT++ simulator with this more realistic one. This component is only available in \emph{integrated-mode}. 

\section{Results and Discussions} \label{results_and_discussions}

This section presents a demonstration of the benefits of an iterative and empirical approach to algorithm development in the context of the Internet of Flying Things. It is aided by GrADyS-SIM NextGen which serves here as a tool for creating simulated environments and easily translating code between them. The demonstration presents a hypothetical scenario which is solved by a distributed algorithm implemented as a protocol in the framework. This protocol will then be tested and iterated as more realistic conditions are added. It is not meant as a revolutionary solution to a real world problem, but as a toy problem where the development of the algorithm itself is more important than the solution it proposes. The scenario will be better explained below.

It is a data-collection scenario set in some remote location deprived of any network infrastructure. A set of stationary sensors has been distributed in arbitrary known locations in the location of interest. These sensors are constantly collecting data from their immediate environment, and that information needs to be collected and processed. The sensors have limited communication range. The remote nature of the location and lack of infrastructure makes remote collection impossible, and the frequency with which data is generated makes manual collection impractical. Some quad-copter UAVs are available. These vehicles are capable of communication with the sensors and each other, and of autonomous flight. A ground station (GS) has been set up with short-range communication equipment capable of talking to the UAVs. 

As for the solution, the UAVs serving as mobile nodes will be employed in collecting data from the sensors and bringing it to the GS. They will fly above the sensors, communicating with them to retrieve the data collected and then return to the GS, delivering the data. The real challenge in this scenario is coordinating UAV movement efficiently to maximize their usefulness. This coordination will happen through communication and will follow some protocol.

The protocol chosen for the solution puts UAVs in a common waypoint mission. This mission starts at the ground station and travels above every sensor in the field. When reaching the end of the mission, the UAV will be on top of a sensor and will need to turn around and complete the mission in reverse in order to come back to the start. While performing this mission, the mobile node will constantly advertise its presence with a heartbeat message. Sensors can receive this messages and will respond with a message containing data collected from its proximity. The mobile node will listen for these messages and absorb the data when received. UAVs also interact with each other, when two meet they will exchange the data they have stored and end their interaction by flying in opposite directions on the mission, with the UAV that ends up traveling towards the GS carrying all the data from the pair. This interaction is managed through a three-step sequence composed of the heartbeat, which is responded with a pair request that is acknowledged with a pair confirmation. The same UAV to UAV interaction also happens between UAV and the GS. With the GS being stationary it of course can't move away from the UAV so it always gets all the data from it and the UAV will end up travelling away from it. This protocol is called ZigZag, and its objective is collaboratively relaying sensor data to the GS.

It is also necessary to describe how the scenario will be represented in the simulations. The sensors are represented by stationary nodes, as is the GS. The UAVs are represented as mobile nodes. Three different protocols were developed, one for each of these node types. The message interactions described in the protocol's specification were implemented, and a simulation scenario was constructed. The GS was placed, and all mobile nodes were initialized in that same position. Sensors were placed in a line in 300 meter intervals away from the ground station. A line was chosen as the sensor's dispersal pattern in the field was deemed irrelevant for this simple experiment. Three configurations of this scenario were built, one with 5 sensors and two UAVs, another with 15 sensors and 7 UAVs and a last one with 25 sensors and 12 UAVs. The number of UAVs was intentionally kept lower than the number of sensors, as having an excessive number of UAVs changes the nature of the problem to something akin to a drone swarm. The information the sensors record from their proximity is being abstracted away as it is not relevant to the scenario, which focuses on the transportation of this information and not on its nature. Every time a UAV receives data from a sensor, it will simply increment a counter, representing the information collected. This information will be exchanged with other UAVs and the ground station.

The scenario was constructed in prototype-mode using the python simulator, and this was the environment used for the initial creation of the protocol. A network communication range of 50 meters was used in this test. An iterative and empirical approach was taken, as pieces of the protocol were slowly added and tested with the aid of the simulation. Initial simulations showed the mobile nodes getting stuck in interaction loops where a new interaction starts after the end of a previous one, before the nodes could fly out of communication range from each other. A timeout strategy was introduced, mobile nodes will ignore communication from others for a couple of seconds after finishing an interaction. Another observation was that with all UAVs starting their mission at the same time, they would stay very close to each other during the simulation's course. A change was made to start them in a staggered manner, with UAVs being released one by one.

After the initial set of changes, a version deemed stable enough was then set up to be tested in integrated-mode using the integration between the protocol interface and OMNeT++ and its component library INET. An IEEE 802.11 compliant radio was used in all nodes, and their transmission power was tuned to approximate the 50-meter range chosen in python. Additionally, background noise was added to the network to simulate the noise and interference one would expect in a real world scenario. The same scenario and three configurations were set up in this new environment. 

Initial results were confusing as no data was being collected in the ground station. Further analysis pointed out that the cause was a flaw in the protocol's behavior that could not be captured by the simpler python simulator. The UAV and ground station heartbeats were sent in regular intervals, which were the same for every UAV. This meant heartbeats were being sent at the exact same time for all nodes. In prototype-mode this wasn't an issue as network conditions are not taken into account when evaluating if a transmission was successfully, only the distance between the recipient and sender. The OMNeT++/INET simulation stack take into account several factors when handling message transmission, including interference. The fact that all the UAVs and the ground station always sent messages at the same time meant that messages always collided and interfered with each other, not being delivered. A short random offset was added to each mobile node and the ground station to prevent further collisions. This fixed the issue, as demonstrated by later simulation runs. 

This serves as an example of how a more realistic simulation scenario can help with the development of more robust protocols. A more realistic simulation can introduce problems like unexpected message timings arising from transmission delays, loss of packets, interference, transmission failures due to obstacles and many more. These problems are not limited to networking, if we were to use a more complex mobility model we would observe problems from node reaction time, imprecision in positioning and others. When exposing an algorithm to these problems, the effects on its logic can be understood and accounted for. 

After a robust enough version of the protocol was finished, massive simulations were run to collect metrics and evaluate its performance. These simulations were run both in prototype-mode and in integrated-mode, although integrated-mode is much more appropriate for this use case. The integration with the more realistic and proven simulator means the results produced are more reliable. Prototype-mode is more useful for iterative development, running shorter simulations and validating a protocol visually and through code assertions. Either way, the evaluation simulations were run in prototype-mode to serve as a point of comparison for this work.

\begin{figure*}[h]
    \centering
    \includegraphics[width=0.95\linewidth]{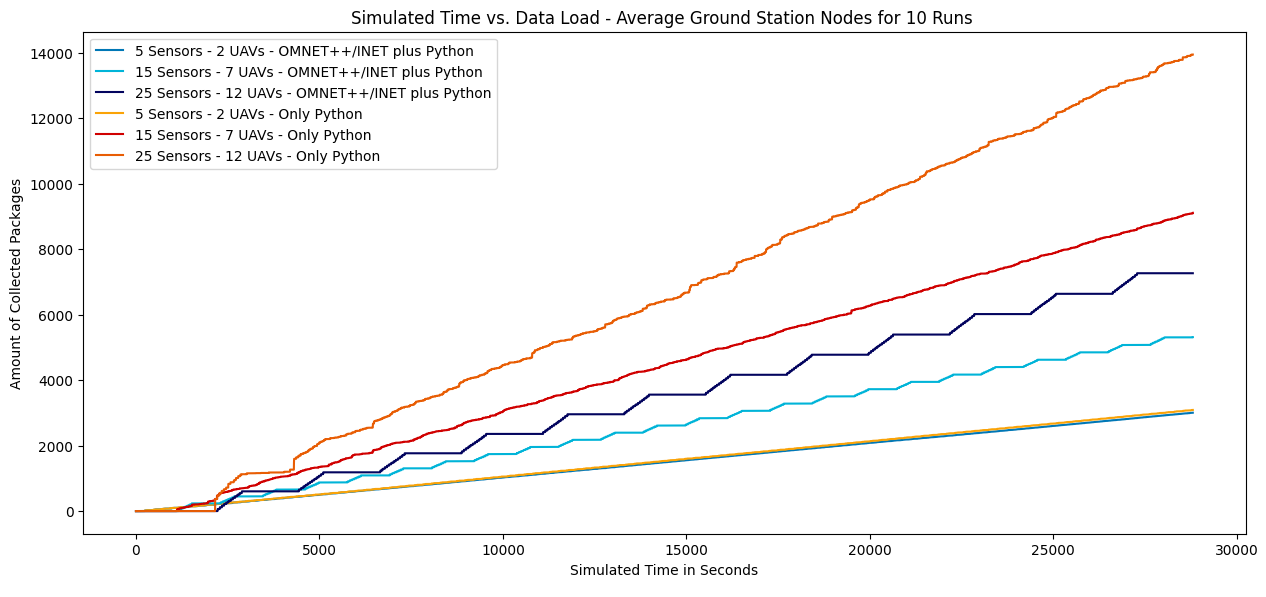}
    \caption{Simulated Time vs. Collected Packages -- Average Ground Station data with 10 Runs}
    \label{fig:dataCollection}
\end{figure*}

The comparison between simulated time and data load for mobile nodes is of significant interest as it helps to understand the protocols' performance under different conditions as well as sheds light on potential trade-offs between simulation realism and simplicity. In Figure \ref{fig:dataCollection}, the plot illustrates the comparison between the three different configurations for both the prototyping and integrated execution of the ZigZag protocol. In the X Axis the figure shows the simulation time and on the Y axis the data count in the ground station at that simulation time. This metric measures the algorithm's effectiveness in transporting data to the ground station. The three configurations for the two simulators are shown as different colored curves. Each configuration was run 10 times and an average of the runs was taken.

The figure shows a big difference in data collected when comparing the same configuration in integrated and prototype modes, with prototype-mode collecting more data. This makes sense as the network simulation of the python simulator is primitive compared to OMNeT++ and INET's modules. They more accurately simulate the loss of packets based on the relative distance between the communicating nodes and take into account the effects of background noise. This observed difference again illustrates the importance of using a more realistic environment. In this particular case, the loss of packets didn't drastically affect the functionality of the protocol, but it is easy to imagine algorithms that are more sensitive to these types of failures. 

\begin{figure*}[h]
    \centering
    \includegraphics[width=0.95\linewidth]{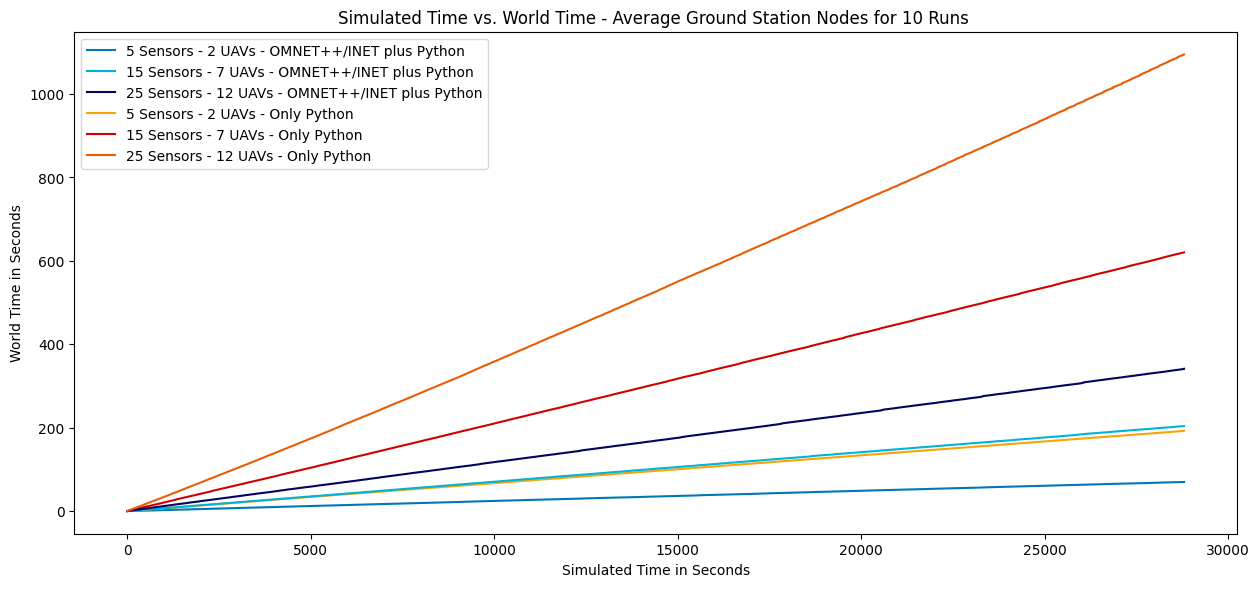}
    \caption{Simulated Time vs. World Time -- Average Ground station data with 10 Runs}
    \label{fig:executionTime}
\end{figure*}

Another metric collected was the simulated time and world time ratio, as seen in figure \ref{fig:executionTime}. This effectively shows how fast the simulator is when running unregulated, as fast as it can go. The X axis shows simulation time and the Y axis shows the real world time it took to get to that simulation time. The steepness of the curve measures how fast the scenario ran, with steeper curves meaning slower times. Each configuration was run 10 times and an average is shown.

The results show that in this metric, integrated mode is dominant. The shortcomings of the python simulator in this can be attributed to a lack of focus on optimization and the usage of Python, which is an interpreted language that is known to pale in comparison to C++ in terms of performance. This doesn't detract from the python simulator's usefulness because, as mentioned previously, running big simulation campaigns is not its focus. Fast iterative development relies on short simulations, not on longer ones. Not pictured in figure \ref{fig:executionTime} are other time wastes that make using OMNeT++ for this kind of development cumbersome. 

First, it relies on a modified Eclipse IDE made for the project. This removes user agency in choosing their own development software and limits them to this IDE, which is very feature rich but slow. Features like syntax highlighting and auto-completing variable names are slow, sometimes temporarily freezing the IDE. Another disadvantage is that there is an additional compilation step before running a simulation. Compilation times ranges from several seconds to a couple of minutes, depending on the amount of files changed relative to previous compilations.

Using an analogy, the speed comparison is akin to a race between a man on foot and a biker, with the biker starting lying on the ground. The racer will get the early advantage, but of course the biker will later pass him. The winner depends on the size of the race, and for iterative development the race doesn't need to be long. Integrated-mode is still the de facto standard for collecting performance metrics, as those depend on longer simulations and benefit from a higher level of realism.

\section{Conclusion}
In conclusion, the proposed approach of iteratively developing distributed algorithms by abstracting away complexity and gradually introducing it, switching simulation environments as that happens, has been empirically validated through the creation and demonstration of the GrADyS-SIM NextGen framework. It aims to address key challenges in the development and testing of distributed algorithms for autonomous vehicles in dynamic, networked environments. It streamlines the intricate processes of creating and evaluating distributed algorithms by offering an environment-agnostic interface for protocol implementation and support for multiple execution environments.

The framework's architecture, as illustrated in Figure \ref{fig:architecture}, is designed with a focus on modularity and flexibility. The protocol module establishes a standardized interface for protocol implementation, ensuring that the same code can run seamlessly across different simulated environments. 

The introduction of \emph{prototype-mode} provides users with a lightweight and accessible simulation environment. With a simplified setup and a focus on rapid prototyping. The \emph{integrated-mode} leverages the power of OMNeT++, a widely used event-based network simulator, to provide a more detailed and accurate representation of network behaviors. The seamless transition between \emph{prototype-mode} and \emph{integrated-mode} ensures that the developed protocols in the initial stages of development can very easily be used in a realistic network simulation.

The experiments presented in section \ref{results_and_discussions} shows the tangible benefits of the proposed approach. During the development of the algorithm, the gradual introduction of complexity made many inadequacies in the logic apparent, without sacrificing the developer experience by introducing every complexity at once.

The next natural step in the development of the framework is supporting real-world scenarios. The foundational for this to be possible has already been laid. All that is left is selecting a hardware testbed and implementing the first integration between it and the framework. This would close the development cycle from inception, to prototyping, to validation and finally to experimentation and deployment.

\section*{Acknowledgments}
This study was financed in part by AFOSR grant FA9550-23-1-0136.

\bibliographystyle{unsrt}  
\bibliography{references}
\end{document}